\documentclass[twocolumn,aps,amsmath,amssymb,prd,nofootinbib]{revtex4}
\usepackage{epsfig}
\usepackage{wasysym}

\def\al{\alpha}

\def\de{\delta}
\def\ep{\epsilon}

\def\et{\eta}
\def\th{\theta}

\def\la{\lambda}

\def\ta{\tau}

\def\ph{\phi}

\def\Ph{\Phi}
\def\Ps{\Psi}
\def\Om{\Omega}

\def\mn{{\mu\nu}}
\def\prt{\partial}

\def\pt#1{\phantom{#1}}

\def\sb{\overline{s}}
\def\ae{(\overline{a}_{\rm eff})}
\def\aes{(\overline{a}_{\rm eff}^S)}
\def\cb{\overline{c}}
\def\cs{\overline{c}^S}

\newcommand{\beq}{\begin{equation}}
\newcommand{\eeq}{\end{equation}}
\newcommand{\bea}{\begin{eqnarray}}
\newcommand{\eea}{\end{eqnarray}}
\newcommand{\rf}[1]{(\ref{#1})}

\def\etal {{\it et al.}}

\def\fr#1#2{{{#1} \over {#2}}}

\def\frac#1#2{{\textstyle{{#1}\over {#2}}}}

\begin{document}

\title{Light-bending tests of Lorentz invariance}
\author{Rhondale Tso\footnote{tsor@my.erau.edu}}
\author{Quentin G. Bailey\footnote{baileyq@erau.edu}}
\affiliation{Physics Department, 
Embry-Riddle Aeronautical University, 
3700 Willow Creek Road, 
Prescott, AZ
86301, USA.}

\date{\today}

\renewcommand{\arraystretch}{1.8}

\begin{abstract}
Classical light bending is investigated for weak gravitational fields 
in the presence of hypothetical local Lorentz violation.  
Using an effective field theory framework
that describes general deviations from local Lorentz invariance, 
we derive a modified deflection angle for light passing 
near a massive body. 
The results include anisotropic effects not present 
for spherical sources in General Relativity
as well as Weak Equivalence Principle violation.
We develop an expression for the relative deflection 
of two distant stars that can be used to analyze
data in past and future solar-system observations.
The measurement sensitivities of such tests to 
coefficients for Lorentz violation are discussed.
\end{abstract}

\maketitle


\section{Introduction}
\label{introduction}

A classic prediction of General Relativity (GR)
is the bending of distant starlight 
by the Sun \cite{ae}. 
This was first confirmed in 1919 by Dyson, 
Eddington, 
and Davidson to agree with Einstein's calculations 
of $1.75''$ at one Solar radii to within $30\%$ accuracy 
\cite{Dys1920}.  
More recent optical measurements 
during solar eclipses have made
only marginal improvements \cite{optical}.
The inclusion of radio astronomy has increased 
the accuracy of light deflection measurements to within $0.02\%$, 
providing additional firm evidence for the 
validity of the light deflection predicted in GR \cite{vlbi}.  
Measurements of the closely-related Shapiro 
time delay have also seen vast improvements recently.
The analysis of two-way radio tracking of 
the Cassini probe matched the predictions of GR
to within parts in $100,000$ \cite{bit}.

Although it is currently the best fundamental 
theory of gravity, 
there remains widespread interest in 
developing more precise tests of GR, 
including improved measurements of the bending of light, 
among others.
These efforts are in part motivated by the intriguing
possibility of finding deviations from GR.
Such deviations could be a signature of a more 
fundamental unified theory of physics
that successfully meshes GR with
quantum theory and the Standard Model of particle physics.

One possible signature that has been sought in 
many sensitive tests are minuscule violations 
of local Lorentz invariance, 
a fundamental tenet of GR \cite{reviews}. 
Theoretical scenarios in which local Lorentz
symmetry could be broken are currently numerous in the literature, 
with early motivation coming from string field theory \cite{ksp}.

In order to investigate violations of local Lorentz invariance, 
it is useful to have a theoretical framework
in which to report measurements.
One systematic framework for studying signals
of Lorentz violation employs effective field theory.
The idea is to incorporate known physics from 
GR and the Standard Model of particle physics, 
into an effective action that also includes 
generic Lorentz-violating terms.
The additional Lorentz-violating terms in the action 
are controlled by coefficients for Lorentz violation, 
which are general coordinate tensor quantities
describing the degree of Lorentz violation 
for each type of interaction 
(gravity, electrodynamics, etc.).
These coefficients can be thought of as 
effectively fixed background fields in spacetime that couple
to curvature and matter fields, 
though their origin can be dynamical \cite{ksp,ssb}.
The framework constructed in this manner 
is known as the Standard-Model Extension (SME) \cite{sme,akgrav},
and has been adopted for numerous tests involving
light, 
matter, 
and gravity \cite{tables}.
Connections between this 
framework and various classic test models for Special Relativity 
are discussed in Refs.\ \cite{km}. 
  
Our focus in this work is on the 
signatures of Lorentz violation for gravitational tests.
In the gravity sector of the SME, 
key signals in a number of experiments and observations 
have been established 
in Refs.\ \cite{qbkgrav, qgrav, tkgrav, qgrav2, bt10}. 
Measurements constraining the coefficients 
for the gravity sector have already begun
using atom-interferometric gravimetry \cite{atom,atom2}, 
lunar laser ranging \cite{bcs2007}, 
and short-range gravity tests \cite{bsl2010}. 
In this paper, 
we analyze one of the fundamental tests of GR, 
the bending of light,
in the effective field theory framework of the SME.
This complements recent work on the related time-delay effect
\cite{qgrav,tkgrav}. 

We begin by deriving a general formula for 
the deflection angle in section \ref{deflection basics}
in terms of an arbitrary post-newtonian metric. 
The post-newtonian metric is described in Sec.\ \ref{post-newtonian metric}.
Assuming a stationary point-like mass
we obtain the deflection angle in a limiting case 
in Sec.\ \ref{light deflection: grazing case}, 
and a more accurate expression in Sec.\ \ref{light deflection: general case}. 
In Section~\ref{solar system tests}, 
we apply these results to light-bending tests in the solar system.
We develop an expression for 
the measurable angle between two stars 
in Sec.\ \ref{relative deflection}.
Details of the relative deflection angle
and methods of analysis are discussed
in Sec.\ \ref{observational analysis}.
We illustrate the observable signals for Lorentz
violation using a near-conjunction example in 
Sec.\ \ref{conjunction example}.
Finally, 
in Sec.\ \ref{summary and estimates},
we summarize the work and estimate the potential
measurement sensitivities of existing and future 
light-bending tests to the coefficients for Lorentz
violation in the gravity sector.
Throughout this work we adopt notation and conventions  
as contained in Refs.\ \cite{qbkgrav,qgrav,tkgrav}.
In particular, 
we work in natural units where $c=1$
and the Minkowski spacetime metric has signature $-+++$.

\section{Theory}
\label{theory}

\subsection{Deflection basics}
\label{deflection basics}

\begin{figure}[h]
\begin{center}
\epsfig{figure=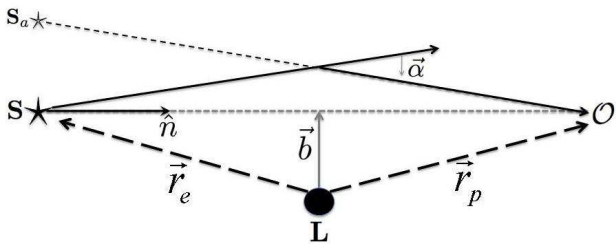,width=0.99\hsize}
\caption{The basic light bending scenario. 
The vector $\vec{r}_p$ extends from the central body $L$ to the observer, 
$\vec{r}_e$ is the vector from the central body to the emitter, 
$\hat{n}$ is in the direction of the unperturbed path, 
and $\vec{b}$ is the impact parameter vector (perpendicular to $\hat{n}$).  
\label{fig0}}
\end{center}
\end{figure}

The deflection $\vec \al$ is the shift
in the direction that light propagates
from a straight line trajectory.
We adopt a simplified gravitational lensing or light-bending scenario 
that involves a source $S$, 
a mass called the lens $L$, 
and an observer $O$.  
The geometric optics limit of electrodynamics 
in curved spacetime is assumed \cite{MTW1973}.
For a point-like lens the apparent source position
observed by $O$ is $S_a$ 
(see Fig.\ \ref{fig0}) \cite{Wambsganss1998,lensing}.  
We assume the lens $L$, 
the source $S$,  
and the observer $O$
are stationary throughout the light ray's propagation.  
The light ray emission is the event with 
coordinates $(t_e, r^j_e)$ and the observation 
of the light ray has coordinates $(t_p,r^j_p)$.

To calculate the deflection we can exploit Fermat's principle:
the null geodesic path from $(t_e,r^j_e)$ to the observer's worldline 
is equivalent to the extremization of the arrival time $t$
on the observer's worldline \cite{SEF1992}. 
For a stationary observer, 
Fermat's principle is equivalent to 
the variational principle:
\beq
\de \int n d\ell = 0.
\label{fermat}
\eeq
Here $n$ is the effective index of refraction of the gravitational field, 
$\ell$ is the euclidean arclength ($d\ell=\sqrt{d\vec x^2}$), 
and they are related by $dt=n d\ell$.
 
The spacetime metric $g_\mn$ is expanded around 
a Minkowski background $\et_\mn$ according to
\beq
g_\mn = \et_\mn + h_\mn.
\label{metricexp}
\eeq
Using the null condition for a light ray ($g_\mn dx^\mu dx^\nu=0$) 
we can evaluate $n$ to leading order in metric fluctuations 
$h_\mn$ as
\beq
n \approx 1+ \frac{1}{2}h_{00} +h_{0j} \fr {dx^j}{d\ell} 
+ \frac 12 h_{jk} \fr {dx^j}{d\ell} \fr {dx^k}{d\ell},
\label{indref}
\eeq
where $dx^j/d\ell$ is the tangent vector to the light path. 
The unit vector $\hat{n}^j$ is the direction 
of the zeroth-order tangent to the light path.
The light trajectory spatial endpoints are $r^j_e$ and $r^j_p$, 
which correspond to the $\ell$ parameter values $l_p$ and $l_e$.  
Referring to Fig.\ \ref{fig0}, 
$l_e=-\vec{r}_e\cdot\hat{n}$ and $l_p=\vec{r}_p\cdot\hat{n}$.
It is useful for later calculations to complete the set $\hat n$
and $\hat b$ with a perpendicular unit vector called $\hat \ta$ 
defined by $\hat{\ta}=\hat{n} \times \hat{b}$.  

If we apply the variational form of Fermat's principle \rf{fermat} 
using the effective index of refraction \rf{indref} we obtain 
equations of motion for the light ray:
\bea
\fr {d^2 x^j}{d\ell^2} &\approx& 
\left(\de^{jk}-\fr {dx^j}{d\ell}\fr {dx^k}{d\ell}\right)
\nonumber\\
&&
\times \left(\prt_k n  
- \fr {d}{d\ell} \Big[ h_{0k}+\fr {dx^l}{d\ell} h_{lk} \Big] 
\right).
\label{eom}
\eea
Note that the terms on the right-hand side
are perpendicular to $dx^j/d\ell$, 
consistent with the definition of euclidean arclength $\ell$.
This equation is equivalent to the geodesic equation 
for light to post-newtonian order $v^2$, 
or PNO(2).

The deflection $\al^j$ follows by integration of Eq.\ \rf{eom}
from the distant source $S$, 
at position $r_e^j$ and $\ell=-l_e$, 
to the observer $O$ at position $r_p^j$ and $\ell=l_p$. 
To PNO(2), 
the resulting deflection is given by the expression
\beq
\al^j = \int\limits_{-l_e}^{l_p} (\prt_j)_\perp n \, d\ell
- \left[\left(h_{0j}+\fr {dx^k}{d\ell} h_{kj} \right)_\perp \right]_{-l_e}^{l_p},
\label{defl:1}
\eeq
where the symbol $\perp$ indicates a
projection perpendicular to $dx^j/d\ell$, 
as in Eq.\ \rf{eom}.
The first integral in \rf{defl:1} is evaluated using the Euclidean
arclength $d\ell$ along the zeroth-order direction of the light ray.  
The second term is to be evaluated at the endpoints and plays a role in the 
result for the case where the observer is near the massive body $L$.

The result in Eq.\ \rf{defl:1} applies to any metric 
that can be expanded around a Minkowski background 
in a post-newtonian series, 
so long as light behaves conventionally 
in the geometric optics limit 
(i.e., light follows a null geodesic).
In the sections that follow,  
we shall apply this result to the post-newtonian
metric that incorporates local Lorentz and WEP 
violations using the SME framework.

\subsection{Post-newtonian metric}
\label{post-newtonian metric}

We focus on the dominant terms in 
the gravitational sector of the SME framework \cite{akgrav}. 
This includes terms augmenting the pure-gravity sector
(terms amending the Einstein-Hilbert action)
as well as Lorentz-violating terms arising from the matter action. 
For the case of linearized gravity, 
the leading corrections to the post-newtonian
metric of GR have been established and are discussed 
in detail in Refs.\ \cite{qbkgrav,tkgrav}.
Using a convenient choice of coordinates and existing constraints
on the vacuum birefringence of light \cite{bire},
we can ignore Lorentz violation in the electromagnetic sector,
and hence assume light propagates normally \cite{tkgrav}.

The Lorentz violation for the pure-gravity sector is controlled 
by $9$ coefficients called $\sb_\mn$.  
For the matter sector, 
the relevant coefficients are $\ae_\mu$ and $\cb_\mn$.
The combined ${\rm PNO(2)}$ metric in harmonic coordinates 
is given by
\bea
g_{00}&=&-1+\big[2+3\sb_{00}+2\cs_{00}+4\frac {\al}{m}\aes_0 \big]U
+\sb_{jk}U^{jk},
\nonumber\\
g_{0j}&=&[\sb_{0j}+\frac {\al}{m}\aes_j ]U
+[\sb_{0k}+\frac {\al}{m} \aes_k] U^{jk},
\nonumber\\
g_{jk}&=&\de_{jk}+[2-\sb_{00}+2 \cs_{00}-2 \frac {\al}{m}\aes_0]\de_{jk}U
\nonumber\\
&& +\de_{jk}\sb_{lm}U^{lm}-\sb_{jl}U^{lk}-\sb_{kl}U^{lj}
\nonumber\\
&& +[2\sb^{00} +2 \frac {\al}{m} \aes_0]U^{jk}.
\label{metric}
\eea
The coefficients for the matter sector, 
$\aes_\mu$ and $\cs_{00}$, 
depend on the structure of the source body $S$
with mass $m$, 
and therefore will differ for distinct source bodies.
Thus $\aes_\mu$ and $\cs_{00}$ indicate the presence
of apparent Weak Equivalence Principle (WEP) violation as well
as Lorentz violation.
Note, 
however, 
that these WEP-violating coefficients 
do not affect the propagation of light rays, 
except through the spacetime metric.
Also, 
as discussed in Ref.\ \cite{tkgrav},
a model dependent scaling $\al$ appears multiplying
the coefficients $\aes_\mu$.
In contrast, 
the coefficients $\sb_\mn$ do not depend
on the nature of the source body and therefore
describe WEP-conserving local Lorentz violation 
for gravity.

Some limiting cases of this metric 
should be noted.
This post-newtonian metric can be viewed as
enlarging and complementing 
the Parametrized Post-Newtonian (PPN) metric \cite{Will2006} 
as discussed, 
for the case of vanishing matter coefficients, 
in Ref. \cite{qbkgrav}.  
Also, 
in the limit that the coefficients $\sb_\mn$, 
$\aes_\mu$, 
and $\cs_{00}$ vanish, 
this result reduces to the post-Newtonian metric of GR.  

For classical light bending scenarios 
we assume the source body $L$ 
to be a point-like mass $M$.  
Furthermore, 
the source body is taken 
as the origin of the coordinate system.  
Thus, 
the dominant contributions to the potentials 
$U$ and $U^{jk}$ depend on the position of the test body $r^j$:
\bea\label{a2}
U&=&\fr {GM}{r}, \\
U^{jk}&=&\fr {GMr^j r^k}{r^3},
\eea
where $G$ is Newton's gravitational constant.

\subsection{Light deflection: grazing case}
\label{light deflection: grazing case}

As a preliminary investigation of the modifications 
to the standard GR light-bending formula, 
we consider the simplified ``grazing'' case. 
In this scenario both the emitter and receiver
are assumed to be very far from the source, 
thus we take $l_p\rightarrow\infty$ and $l_e\rightarrow\infty$.
Using the metric \rf{metric} in the general result \rf{defl:1}
we obtain for this limiting case the deflection
\bea
\al^{j}&=&\fr {-4GM}{b}
\Big( 
[1+\sb_{00}+\cs_{00}+\frac {\al}{M}\aes_0
\nonumber\\
&& \pt{\fr {4GM}{b}} 
+(\sb_{0k}+\frac{\al}{M}\aes_k) \hat{n}^{k}] \hat{b}^{j}
-\sb_{kl}\hat{b}^k \hat{\tau}^l 
\hat{\tau}^j
\Big).\nonumber\\
\label{defl:2}
\eea
The standard GR result is obtained 
in the limit that the coefficients for Lorentz violation vanish.  

To illustrate some of the features of
the result \rf{defl:2} we employ vector field
plots indicating the apparent shift of incoming starlight
as the source appears in front of a background of
initially uniformly distributed stars.
This is similar to methods that have been adopted for the GR signal 
in Refs.\ \cite{cm, kop06}. 
For this simulation we focus on a small patch of sky 
centered on the deflecting body. 
We use $x$ and $y$ coordinates on the patch to 
make a vector field plot of different parts of 
the result \rf{defl:2}.
These coordinates will lie in the plane perpendicular
to the unit vector $\hat n$, 
which we can take to point (approximately) from the 
deflecting body to the observer.\footnote{Strictly speaking, 
$\hat n$ points from the position of each star to the observer, 
but in the ``grazing'' approximation we can ignore the variation 
over the patch of sky.}
  
\begin{figure}[h]
\begin{minipage}{1.67in}
\epsfig{figure=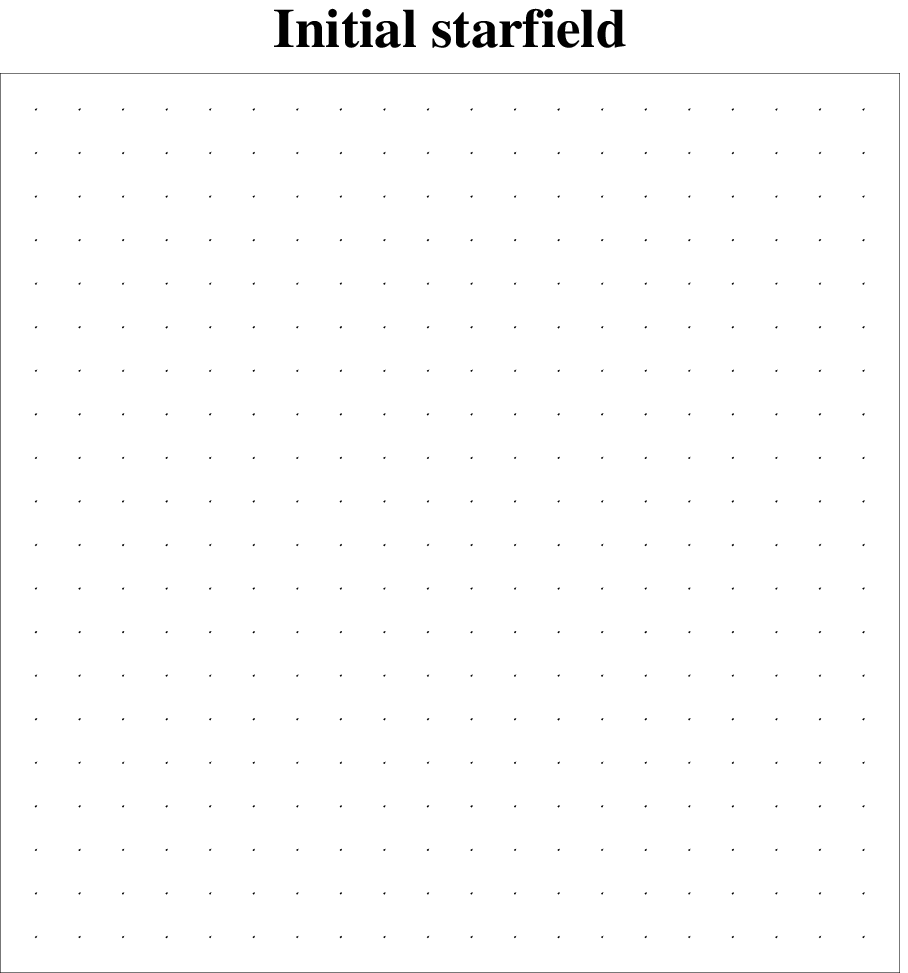,width=1.0\hsize}
\end{minipage}
\hfill 
\begin{minipage}{1.67in}
\epsfig{figure=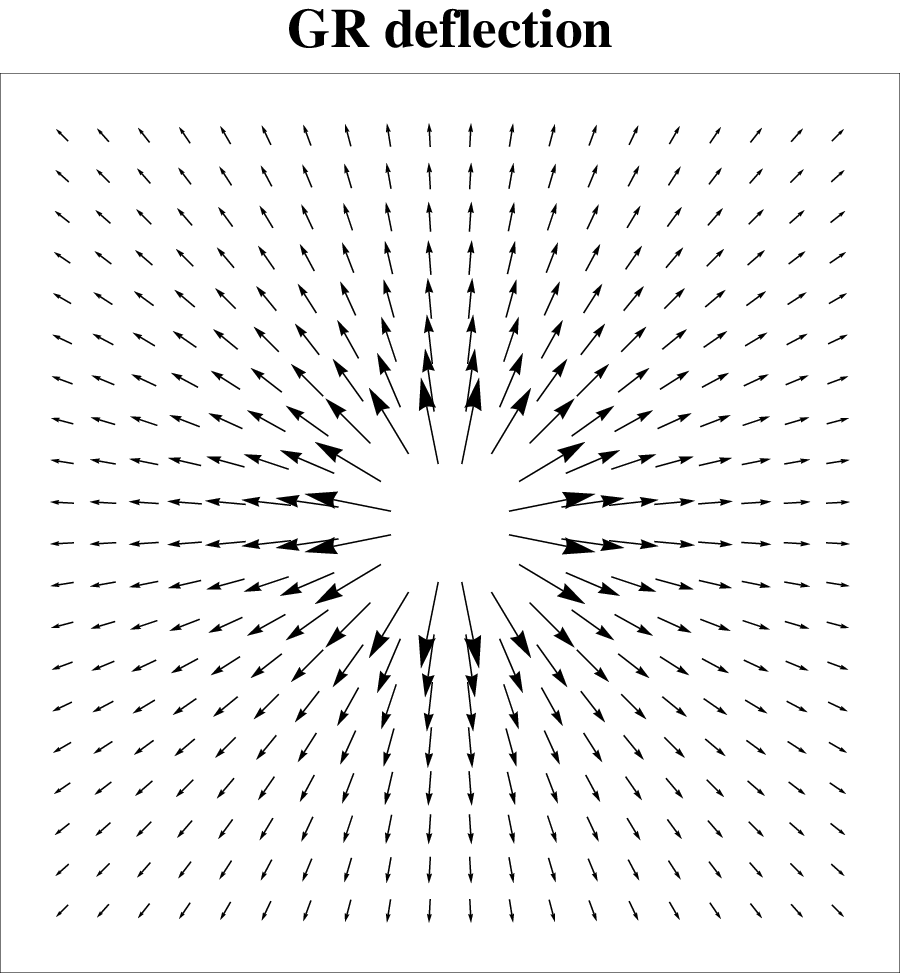,width=1.0\hsize}
\end{minipage}
\caption{Initial uniform star field (left). 
Apparent shift of distant light 
in the standard general relativity case
(right).\label{fig1}}
\vskip4mm
\begin{minipage}{1.67in}
\epsfig{figure=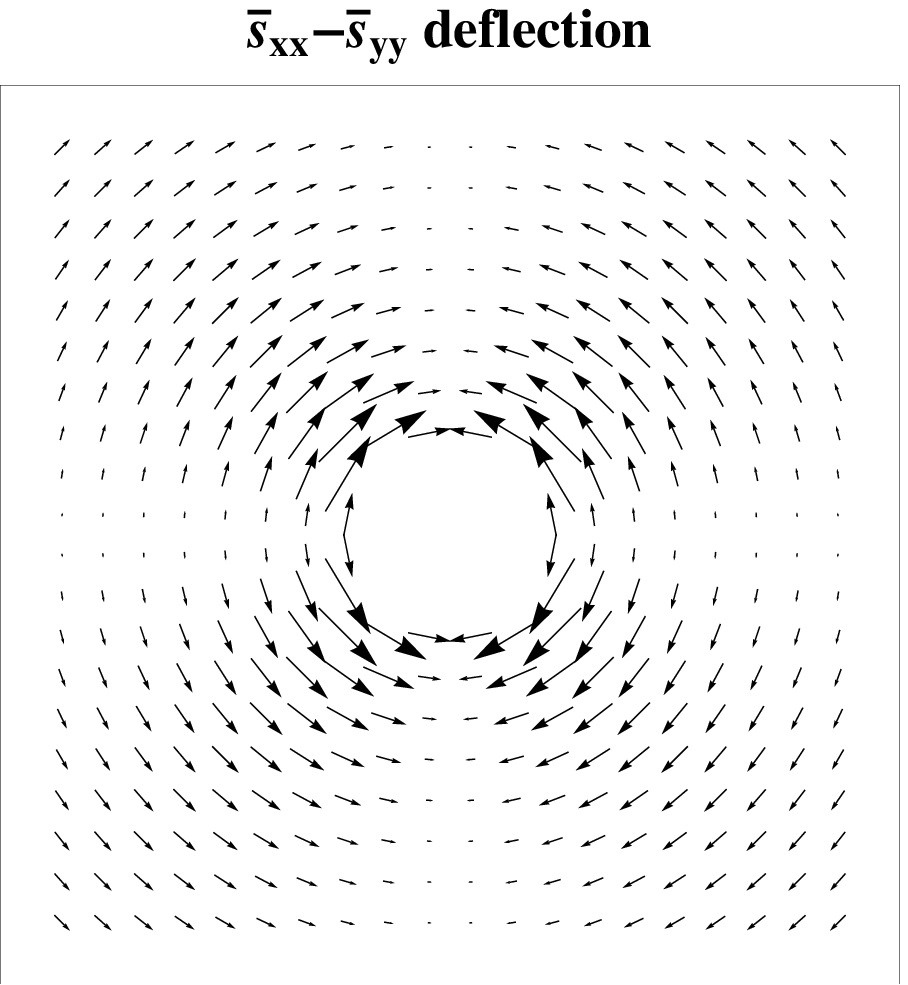,width=1.0\hsize}
\end{minipage}
\hfill 
\begin{minipage}{1.67in}
\epsfig{figure=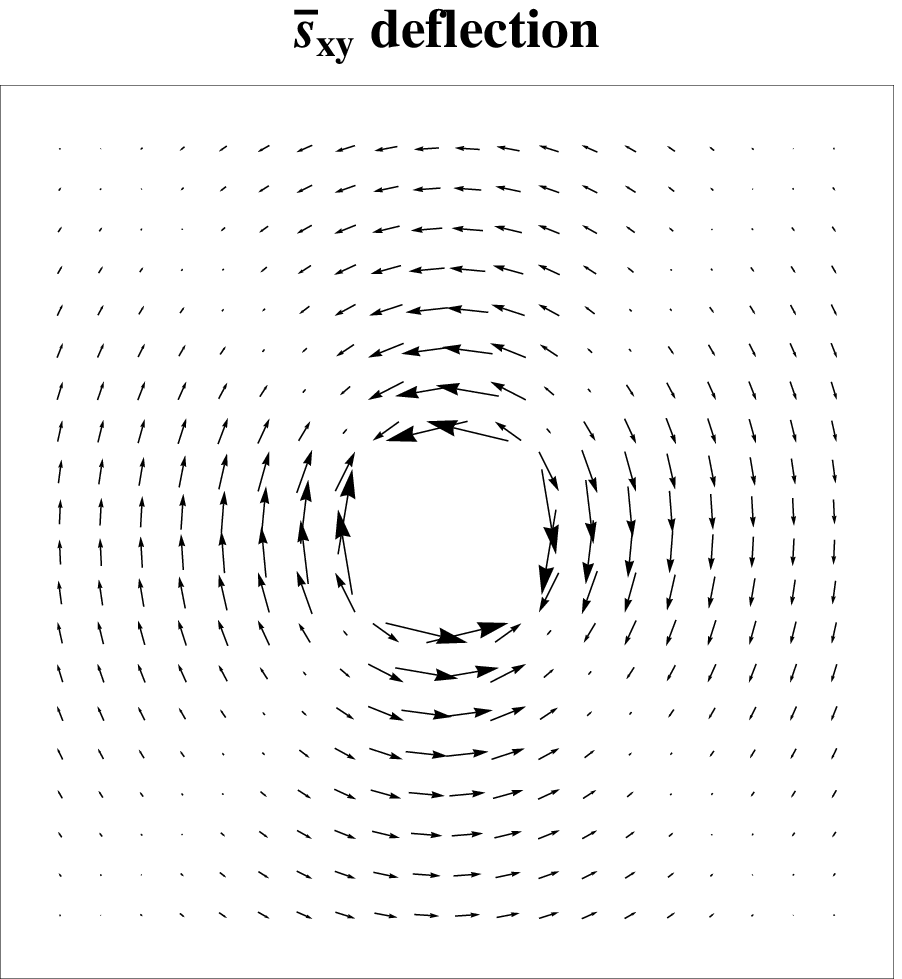,width=1.0\hsize}
\end{minipage}
\caption{Anisotropic apparent shift of star field
due to the $\sb_{xx}-\sb_{yy}$ 
coefficients (left) 
and the $\sb_{xy}$ coefficients (right).
The local $x$ coordinate runs horizontally
and $y$ is vertical.
\label{fig2}}
\end{figure}

The left panel of 
Fig.\ \ref{fig1} shows the 
initial star field in the absence of the source.
For simplicity, 
we assume the distribution of stars is uniform. 
The right panel of Fig.\ \ref{fig1} shows the deflection 
in the GR limit.
In the expression for the deflection \rf{defl:2}, 
the term proportional to $\hat b^j$ 
is scaled by the coefficients for Lorentz violation
$\sb_{00}+\cs_{00}+(\al/M)\aes_0$
and the $\hat{n}$-dependent 
combination $(\sb_{0k}+(\al/M)\aes_k) \hat{n}^{k}$.
Note that, 
due to the appearance of
these coefficients, 
this scaling will change depending 
on the location of the patch of sky considered
and the internal nature of the source.
That this occurs is due to the anisotropy of
the coefficients for Lorentz violation
and their composition dependence.

The last term in Eq.\ \rf{defl:2} points 
in the direction $\hat \ta$, 
orthogonal to $\hat{b}$.
This term is proportional to the combination
of coefficients $\sb_{kl} \hat{b}^k \hat{\tau}^l$.
If we express this combination in terms
of the ($x$, $y$) coordinates on the patch 
of sky considered, 
this can be expanded into the 
two independent combinations 
$\sb_{\,xx}-\sb_{\,yy}$ and $\sb_{\,xy}$. 
The deflections due to these two combinations 
of coefficients are plotted in Fig.\ \ref{fig2}, 
where we have set their values to one for illustration.
These plots indicate that anisotropic 
deflection occurs for nonvanishing 
coefficients $\sb_{jk}$.
In particular, 
this means that a light ray would  
deflect off of an otherwise flat plane defined 
by $\hat{n}$ and $\hat{b}$, 
contrary to the conventional GR point-mass case.
Note that these anisotropic effects can in principle
be distinguished from the GR deflection due to 
higher multipole moments of the deflecting body \cite{zk,kop06} 
by virtue of their $1/b$ dependence. 

It is interesting to compare this result with that obtained 
in Refs.\  \cite{qgrav,tkgrav} for the gravitational time delay 
in the presence of local Lorentz violation.
The vector coefficients $\sb_{0j}$ and $(\al/M) \aes_j$ appear in the 
light bending result \rf{defl:2}, 
while they are absent in the round-trip time delay signal.
Also, 
light-bending observations that involve one orientation
of the observer, 
deflecting body, 
and the distant starlight
could gain access to sets of coefficients distinct from 
those in a dedicated time-delay test with a different orientation
of the relevant bodies.
Note that this feature does not occur in the PPN formalism, 
for which both types of tests access the same parameter
regardless of the underlying orientation.

\subsection{Light deflection: general case}
\label{light deflection: general case}

For applications of \rf{defl:1} to observations, 
the light-grazing approximation previously 
assumed must be reconsidered.  
Here we still assume $l_e\rightarrow \infty$, 
but $l_p$ is treated as a relevant, 
finite term.  
This corresponds to the case where the observer is a finite
distance from the lens $L$, 
while the light source is effectively at spatial infinity.
Referring to Fig.\ \ref{fig0} it will be useful to 
define an angle $\Ph$ between $\vec r_p$ and $\hat n$
(thus $\hat{r}_p \cdot \hat{n}= \cos\Ph$). 

Evaluating the deflection formula \rf{defl:1}
using the metric \rf{metric} 
for the case $l_e\rightarrow \infty$, 
we obtain the deflection,
\bea
\al^j &=& \fr {-GM}{b} \Bigg[  \hat{b}^j 
\bigg( 
2(1+\sb_{00}+ \cs_{00}+\frac {\al}{M}\aes_0
\nonumber\\
&& \pt{+}+(\sb_{0k}+\frac {\al}{M}\aes_k) \hat{n}^k) (1 + \cos \Ph) 
\nonumber\\
&& \pt{+}+(\sb_{00} + \frac {\al}{M}\aes_0-\sb_{kl} \hat{n}^k \hat{n}^l)
\cos \Ph \sin^2 \Ph 
\nonumber\\
&& \pt{+}+2(\sb_{0k}+\frac {\al}{M}\aes_k) \hat{b}^k \sin \Ph
-\sb_{kl} \hat{n}^k \hat{b}^l \sin^3 \Ph 
\bigg)
\nonumber\\
&&+\hat{\ta}^j
\bigg(
2(\sb_{0k}+\frac {\al}{M}\aes_k) \hat{\ta}^k \sin \Ph
+\sb_{kl}\hat{n}^k\hat{\ta}^l \sin^3 \Ph
\nonumber\\
&&
\pt{+\ta^j(}-\sb_{kl}\hat{b}^k \hat{\ta}^l (2+2\cos \Ph +\cos \Ph \sin^2 \Ph)
\bigg)\Bigg]\nonumber\\
\label{defl:3}
\eea
where the substitutions of $l_p/r_p = \cos \Ph$ 
and $b/r_p = \sin \Ph$ have been made.  
Note that in the light-grazing limit 
the $\sin \Ph$ term vanishes and 
the $\cos \Ph$ term approaches unity, 
which results in the previous deflection angle 
presented in \rf{defl:2}.  

The result \rf{defl:3} indicates that additional 
projections of coefficients for Lorentz violation
arise in the more accurate deflection formula.
Also, 
these additional combinations of coefficients appear 
to be distinct from the those that occur in the 
gravitational time delay derived in Refs.\ \cite{qgrav, tkgrav}.
By itself, 
the deflection in Eq.\ \rf{defl:3}
is not directly measurable.
This is because only the apparent 
position of a given source star is actually measured
(at least during a single observation period).
Instead, 
a comparative measurement is needed based
on two or more observations.
In the following section we apply this result to calculate 
the relative deflection of two stars, 
which is a measurable quantity.

\section{Solar System Tests}
\label{solar system tests}

The key observable of interest for typical 
light bending measurements is the relative deflection of two 
(or more) stars. 
We seek here the angle between a ``source'' star
and a ``reference'' star. 
In what follows quantities associated with the reference star will
have $r$ subscripts, 
and those for the source star will have no subscripts. 
The apparent positions of both of these stars will in principle
be gravitationally deflected if the lens is near the line of sight.
By continuous monitoring of the relative position of these two stars, 
one can measure the effects of the deflection \rf{defl:3}.
Furthermore, 
the formula we derive below can in principle 
be applied repeatedly to systems of many stars.

\subsection{Relative deflection}
\label{relative deflection}

\begin{figure}[h]
\begin{center}
\epsfig{figure=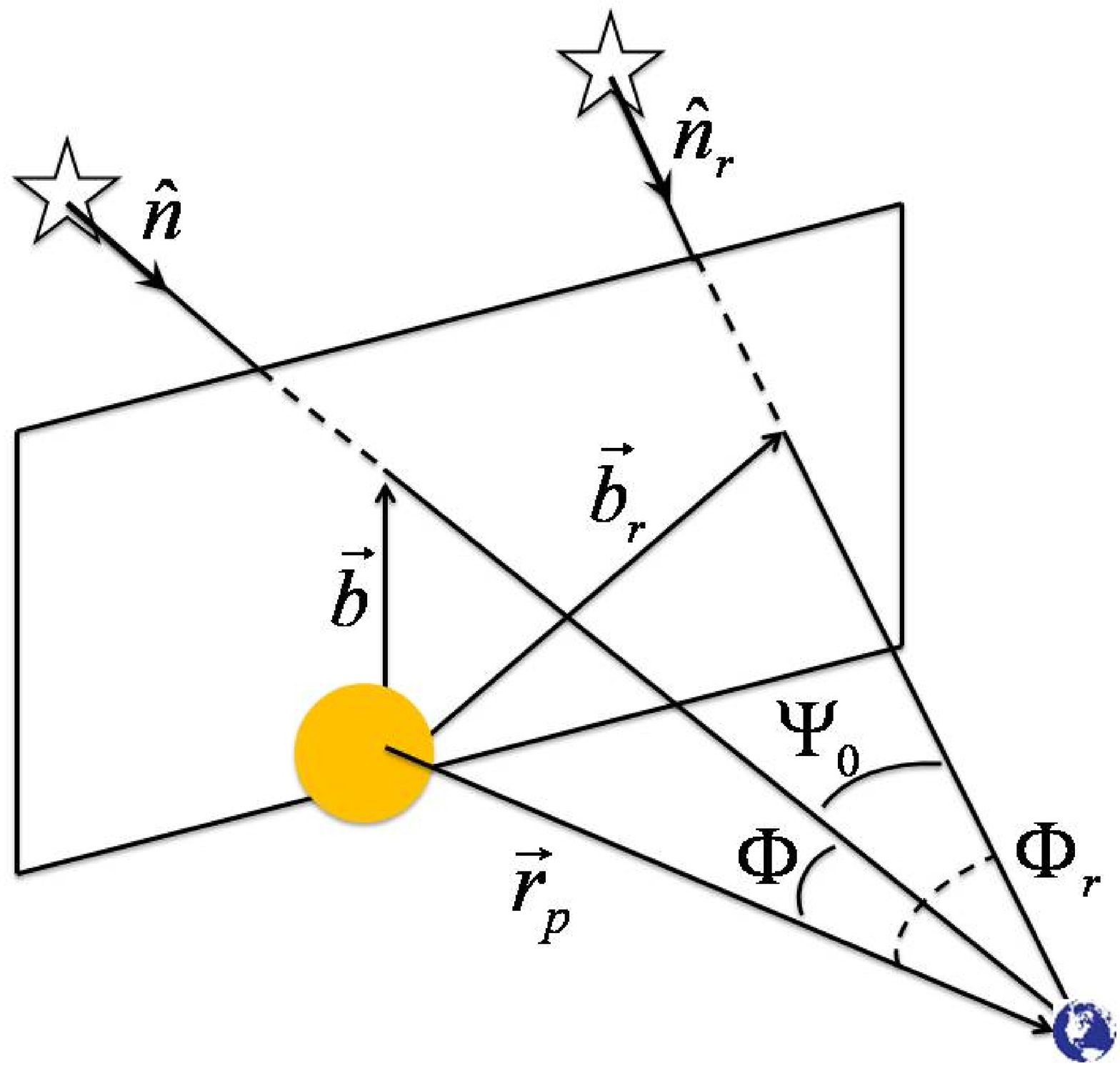,width=1.0\hsize}
\caption{Solar light-bending scenario in the presence of a reference source.  
For the source and reference star, 
we have the impact parameters $\vec{b}$
and $\vec{b}_r$, and directions $\hat{n}$ and $\hat{n}_r$,
respectively. 
The reference plane shown is that 
spanned by $\vec{b}$ and $\vec{b}_r$.
\label{lightbend}}
\end{center}
\end{figure}

To calculate the relative deflection 
we adopt standard methods in the literature \cite{Will2006, gaia2}. 
We begin with a general coordinate invariant expression 
for the angle $\Ps$ between two stars:
\beq
\cos \Ps = 1+\fr {p^\mu (p_r)_\mu}{(u^\nu p_\nu)(u^\la (p_r)_\la)}.
\label{cosps}
\eeq
In this expression $p^\mu$ and $(p_r)^\mu$ are the tangent four-vectors
for the source light ray and the reference light ray, 
respectively.
The four-velocity of the observer measuring the 
relative angle is $u^\mu$.
Evaluating this expression to ${\rm PNO(2)}$, 
and neglecting aberration terms, 
we obtain
\beq
\cos \Ps = \hat{n} \cdot \hat{n}_r
+\hat{n} \cdot (\vec \al_{\rm eff})_r
+\hat{n}_r \cdot \vec \al_{\rm eff},
\label{cosps:2}
\eeq
where $\vec \al_{\rm eff}$ is given by
\beq
\al_{\rm eff}^j = \al^j + \frac 12 (\hat{n}^k h_{kj})_\perp |_{\ell=\ell_p}.
\label{aleff}
\eeq
The deflection $\al^j$ is obtained from Eq.\ \rf{defl:3} and
quantities with the label $r$ are obtained by 
re-labeling all quantities occurring in the expression,
involving the direction of the light ray,
with the subscript $r$ 
(e.g., $\hat{n}\rightarrow\hat{n}_r$,
$\hat{b}\rightarrow\hat{b}_r$, etc. ).
The zeroth order, 
or straight line, 
trajectories from the source and reference stars
are depicted in Fig.\ \ref{lightbend}.

We define an angle $\Ps_0$
that represents the unperturbed angle between 
the two stars:
\beq
\hat{n}_r \cdot \hat{n} = \cos \Ps_0
\label{ps0}
\eeq
Thus $\Ps_0$ is the angle between the two stars
in flat spacetime in the absence of other 
conventional effects such as aberration.
The shift or change in the angle between the two stars 
is defined to be
\beq 
\de \Ps = \Ps -\Ps_0.
\label{delps:1}
\eeq

Since we are working in the post-newtonian 
approximation it suffices 
to assume $\de \Ps$ is small so that 
$\cos \Ps \approx \cos \Ps_0 - \de \Ps \sin \Ps_0$.
Using this approximation, 
the previous result \rf{defl:3}, 
and the metric in \rf{metric},
we obtain
\bea
\sin \Ps_0 \de \Ps &\approx& 
\fr {GM}{b} 
\left( \hat{n}_r \cdot \hat{b} [2(1+\cos \Ph)+B]
+\hat{n}_r \cdot \hat{\ta} T \right)\nonumber\\
&&
\hskip-10pt
+\fr {GM}{b_r} 
\left( \hat{n} \cdot \hat{b}_r [2(1+\cos \Ph_r)+B_r]
+\hat{n} \cdot \hat{\ta}_r T_r \right)\nonumber\\
\label{delps:2}
\eea
In \rf{delps:2} the terms $B$, $T$, $B_r$ and $T_r$ 
are proportional to combinations of the 
coefficients for Lorentz violation $\sb_\mn$,
$\cs_{00}$, and $\aes_\mu$.
Explicitly, 
they are given by
\bea
B&=&
2[\sb_{00}+ \cs_{00}+\frac {\al}{M}\aes_0
\nonumber\\
&&\pt{2} +(\sb_{0k}+\frac {\al}{M}\aes_k) \hat{n}^k] (1 + \cos \Ph) 
\nonumber\\
&& +\frac 12 (\sb_{kl} \hat{b}^k \hat{b}^l
-\sb_{kl} \hat{n}^k \hat{n}^l)\cos \Ph \sin^2 \Ph 
\nonumber\\
&& +2(\sb_{0k}+\frac {\al}{M}\aes_k) \hat{b}^k \sin \Ph
\nonumber\\
&&
\pt{2}-\sb_{kl} \hat{n}^k \hat{b}^l \sin \Ph (\sin^2 \Ph-\frac 12)
\label{B}\\
B_r&=& 
2[\sb_{00}+ \cs_{00}+\frac {\al}{M}\aes_0
\nonumber\\
&& \pt{2}+(\sb_{0k}+\frac {\al}{M}\aes_k) \hat{n}_r^k] (1 + \cos \Ph_r) 
\nonumber\\
&& +\frac 12 (\sb_{kl} \hat{b}_r^k \hat{b}_r^l
-\sb_{kl} \hat{n}_r^k \hat{n}_r^l)\cos \Ph_r \sin^2 \Ph_r 
\nonumber\\
&& +2(\sb_{0k}+\frac {\al}{M}\aes_k) \hat{b}_r^k \sin \Ph_r
\nonumber\\
&&
\pt{2}-\sb_{kl} \hat{n}_r^k \hat{b}_r^l \sin \Ph_r (\sin^2 \Ph_r-\frac 12)
\label{Br}\\
T&=&
2(\sb_{0k}+\frac {\al}{M}\aes_k) \hat{\ta}^k \sin \Ph
\nonumber\\
&&
+\frac 12 \sb_{kl}\hat{n}^k\hat{\ta}^l \sin \Ph (1+\sin^2 \Ph)
\nonumber\\
&&
-\sb_{kl}\hat{b}^k \hat{\ta}^l (2+2\cos \Ph +\frac 12 \cos \Ph \sin^2 \Ph)
\label{T}\\
T_r&=& 2(\sb_{0k}+\frac {\al}{M}\aes_k) \hat{\ta}_r^k \sin \Ph_r
\nonumber\\
&&
+\frac 12 \sb_{kl}\hat{n}_r^k\hat{\ta}_r^l \sin \Ph_r (1+\sin^2 \Ph_r)
\nonumber\\
&&
-\sb_{kl}\hat{b}_r^k \hat{\ta}_r^l (2+2\cos \Ph_r 
+\frac 12 \cos \Ph_r \sin^2 \Ph_r)
\label{Tr}
\eea
The usual result from general relativity is obtained 
for vanishing coefficients for Lorentz violation, 
which follows here in the limit $B=T=B_r=T_r=0$. 

The result in Eq.\ \rf{delps:2} exhibits several interesting 
features that do not occur in the point-mass limit of GR. 
First, 
the rotational scalar combinations involving
$\sb_{00}$, 
$\cs_{00}$, 
and $\aes_0$ scale the usual GR result.
However, 
due to the composition dependence of $\cs_{00}$ and $\aes_0$, 
this scaling will vary with the central body (or Lens) producing the 
gravitational deflection.
Secondly, 
the anisotropic coefficients $\sb_{0j}$, 
$\aes_j$, 
and $\sb_{jk}$ occur in the deflection result
such that they are projected along directions
associated with the position of the source and reference star
($\hat{n}$, $\hat{n}_r$, etc.).\
This implies that observations made either with significantly 
changing star positions, 
or made with different sets of stars at different locations 
in the sky, 
will experience different deflections.
This effect is independent of the conventional $\Ph$, 
$\Ph_r$, and $\Ps_0$ dependence of the deflection.
Thirdly, 
deflection occurs in the directions $\hat{\ta}$ and $\hat{\ta}_r$ 
perpendicular to $\hat{n}$ and $\hat{n}_r$, 
as already illustrated in Figs.\ \ref{fig2}.
The anisotropic effects imply that observations made over time or with many stars would yield access to different combinations of coefficients for Lorentz violation, 
thus increasing the potential ``parameter space'' of possible types of 
Lorentz violations to which light deflection observations could be sensitive.

\subsection{Observational analysis}
\label{observational analysis}

The form of Eq.\ \rf{delps:2} indicates that it 
depends on a number of parameters which can 
be related to specific measurable astronomical quantities, 
in addition to its dependence on coefficients for Lorentz violation.
To illustrate this we describe in this section 
how analysis might proceed.
Our post-newtonian coordinate system is taken to coincide
with the standard Sun-centered celestial equatorial 
coordinate system adopted for many studies of Lorentz violation. 
The spatial coordinates are denoted by $X$, 
$Y$, 
and $Z$ with $X$ pointing in the direction of the Sun
at the vernal equinox, 
and $Z$ is aligned with the Earth's rotation axis.
The time coordinate is $T$ and is typically defined
so that $T=0$ at the $2000$ vernal equinox.
Details of the Sun-centered celestial equatorial coordinate system 
can be found in Refs.\ \cite{scf,tables}. 
In particular, 
the reader is referred to a depiction
of this coordinate system in Fig.\ $1$ of Ref.\ \cite{tables}.

In general, 
for data analysis, 
one can seek to express the relative deflection \rf{delps:2}
such that its functional form is 
\beq
\de \Ps = \de \Ps (\th, \th_r, \ph, \ph_r, T, 
\sb_\mn, \aes_\mu, \cs_{TT},...).
\label{delps:dep}
\eeq
The first four parameters $( \th, \phi)$  
and $(\th_r, \phi_r )$ determine,
respectively, 
the direction of $\hat{n}$ and $\hat{n}_r$ relative 
to the Sun-centered frame.
Explicitly, 
the unit vectors take the form
\bea
\hat{n} &=& -(\sin \th \cos \ph, \sin \th \sin \ph, \cos \th), \nonumber\\
\hat{n}_r &=&-(\sin \th_r \cos \ph_r, \sin \th_r \sin \ph_r, \cos \th_r).
\label{nhats}
\eea
Specifically,
($\th$, $\ph$) and ($\th_r$, $\ph_r$) are taken as
co-latitude and right ascension on the celestial sphere
for the source star and reference star, 
respectively.
In terms of these angles,
the angle $\Ps_0$ can be determined using Eq.\ \rf{ps0}
and the angles $\Ph$ can $\Ph_r$ can be determined using
$\hat{r}_p \cdot \hat{n}= \cos\Ph$ and 
$\hat{r}_p \cdot \hat{n}_r= \cos\Ph_r$.\footnote{In principle, 
the coordinates of the source star and reference star are 
also affected by gravitational deflection.  
However, 
the deflection expression \rf{delps:dep} is already
expressed at ${\rm PNO(2)}$, 
and so within this expression, 
initially measured or ``catalog'' values can be used.}

The time parameter $T$ appears in part due to the 
observer's motion in the Sun-centered frame.
To sufficient accuracy in any of the Lorentz-violating terms
of \rf{delps:dep} we can assume that the motion of the observer
is explicitly known.
For example, 
for the Earth we can assume a circular orbit and use
\beq
\hat{r}_p = (-\cos \Om T, -\cos \et \sin \Om T, -\sin \et \sin \Om T),
\label{earth}
\eeq
where $\et$ is the inclination of the ecliptic to the equatorial plane 
and $\Om$ is the Earth's orbital frequency.

The last set of parameters $\sb_\mn$, $\aes_\mu$,
and  $\cs_{TT}$ are the coefficients for Lorentz violation
expressed in the Sun-centered frame coordinates.  
In addition to the dot products 
($\hat{n}_r \cdot \hat{b}$, $\hat{n}_r \cdot \hat{\ta}$, etc.), 
the terms occurring in the deflection formula \rf{delps:2}
contain projections of the coefficients along 
the six unit vectors for the source and reference source
$\{ \hat{n}, \hat{b}, \hat{\ta}, \hat{n}_r, \hat{b}_r, \hat{\ta}_r \}$.
To capture the orientation dependence of the coefficients 
we must express all of these unit vectors in the Sun-centered frame.
This can be accomplished using \rf{nhats} and \rf{earth}.
The impact parameter vector is defined by 
$\vec b = \vec{r}_p-\hat{n} (\hat{n} \cdot \vec{r}_p)$
(see Fig.\ \ref{fig0}).
Using this, 
the unit vectors $\hat{b}$ and $\hat{b}_r$ can be written as
\bea
\hat{b} &=& \fr {\hat{r}_p - \hat{n} \cos \Ph}{\sin \Ph}, 
\nonumber\\
\hat{b}_r &=& \fr {\hat{r}_p - \hat{n}_r \cos \Ph_r}{\sin \Ph_r}. 
\label{bhats}
\eea
From this expression and \rf{nhats} the unit vectors
$\hat{\ta}$ and $\hat{\ta}_r$ can be constructed 
using the cross products
\bea
\hat{\ta} &=& \hat{n} \times \hat{b}, 
\nonumber\\
\hat{\ta}_r &=& \hat{n}_r \times \hat{b}_r. 
\label{tauhats}
\eea

The full set of unit vectors allows us to express
projections of the coefficients in terms of Sun-centered
frame quantities.
For example, 
if only $\sb_{XZ}$ happens to be nonzero, 
we obtain for the projection $\sb_{JK} \hat{\ta}^J \hat{b}^K$, 
\bea
\sb_{JK} \hat{\ta}^J \hat{b}^K &=& \sb_{XZ} 
(\hat{\ta}^X \hat{b}^Z+\hat{\ta}^Z \hat{b}^X)
\nonumber\\
 &=& \csc ^2 \Ph \sb_{XZ} 
[\sin \th 
\nonumber\\
&& \times
   (\sin \ph \cos \Om T-\cos \eta \cos \ph \sin \Om T ) 
\nonumber\\
&& \times
   (\cos \Om T -\sin \th \cos \Ph \cos \ph)
\nonumber\\
&&
 +\sin \Om T 
   (\sin \et \sin \th \sin \ph-\cos \et \cos \th) 
\nonumber\\
&& \times
   (\cos \th \cos \Ph-\sin \et \sin \Om T) ],
\label{stb}
\eea
where we assumed the case of an Earth observer 
and made use of the formula \rf{earth} for $\hat{r}_p$.

The full expression of the deflection angle \rf{delps:2},
in terms of the parameters indicated
in \rf{delps:dep}, 
can be obtained using the methods just described.
This calculation is lengthy,
and so we do not include it here, 
but it should be straightforward to include in a suitable data analysis code.
To give a flavor of the leading Lorentz-violating effects in light bending,
we determine the approximate form of \rf{delps:2}
for a special case in the next subsection.

Finally we note that we 
have included ellipses in expression \rf{delps:dep} to indicate
that we have not discussed a number of other astronomical 
parameters that may be important to a rigorous analysis, 
such as the effects of parallax and proper motion.
Furthermore, 
we have neglected aberration effects in the result \rf{delps:2} that 
depend on the Earth's or observer's velocity. 
We find that one effect of these terms 
is that they multiply the coefficients for 
Lorentz violation in \rf{delps:2} by higher powers of velocity, 
resulting in ${\rm PNO(3)}$ effects.
In addition, 
terms proportional to $\vec v$ or its square
that arise for aberration in the conventional case \cite{soffel}, 
can implicitly depend on the coefficients for Lorentz
violation through orbital dynamics. 
The effects of Lorentz violation on orbits can 
in principle be incorporated using the 
equations of motion derived in Refs.\ \cite{qbkgrav,tkgrav}.

\subsection{Conjunction example}
\label{conjunction example}

To elucidate the Lorentz-violating effects
in the main result \rf{delps:2}, 
we work with a special case of the general SME deflection result
that involves only one set of coefficients for Lorentz violation.
We set to zero all coefficients save those
contained in $\sb_{JK}$.
We also ignore any contributions from $\sb_{JJ}=\sb_{TT}$
that scale the GR deflection. 

We focus in this example on an Earth observer 
during times near the summer solstice when the Earth 
lies below the negative $Y$ axis in the Sun-centered frame
(the summer solstice occurs when $\Om T=\pi/2$).
To exploit the peak behavior of the deflection result
\rf{delps:2}, 
we suppose that the source star is located on the celestial
sphere so that its light 
just grazes the Sun on its way to earth.
The reference star is taken to be a considerable angular distance
away from the source star so that its gravitational 
deflection can be ignored to a good approximation.
For simplicity, 
both stars are assumed to have $\ph=\pi/2=\ph_r$.

In this near-conjunction scenario, 
the formula for the unit vector in 
the direction of the Earth observer is given by
\rf{earth} while the unit vector
for the source star is given by
\beq
\hat{n} = -(0, \cos (\et +\ep), \sin (\et+\ep)), 
\label{nhat:conj}
\eeq
where $\ep$ is a small angle indicative
of the near-conjunction approximation adopted. 
The other needed unit vectors $\hat{b}$ and $\hat{\ta}$ 
can be obtained from the results in the previous subsection.

If we focus on only the dominant terms in 
this scenario, 
the result \rf{delps:2} simplifies considerably.
As previously stated, 
the second term in \rf{delps:2} 
proportional to $GM/b_r$ that involves
the reference star quantities can be neglected
since $b<<b_r$.
We can also discard any terms with one or more powers
of $\sin \Ph$, 
since they will be suppressed in this limit.
Finally, 
using the small angle approximation for $\ep$
can further simplify the expression.
 
If $t$ is the time measured from the summer solstice, 
the GR portion of the deflection becomes
\beq
\de \Ps_{\rm GR} \approx  -2\fr {GM_{\odot}}{R} 
\fr {\sin \ep}{(1-\cos \Om t \cos \ep)}
\label{delps:GR}
\eeq
where $R$ is the earth's orbital radius and $M_{\odot}$
is the Sun's mass.
The portion of the deflection controlled by
the coefficients for Lorentz violation $\sb_{JK}$
stems from the $\sb_{JK}\hat{\ta}^J \hat{b}^K$ term
in Eq.\ \rf{delps:2}.
It is given by
\bea
\de \Ps_{\rm LV} &\approx&  -\fr {GM_{\odot}}{R} 
\fr {\sin \Om t}{(1-\cos \Om t \cos \ep)^2}
\nonumber\\
&&
\times (\sb_A \sin \Om t \cos \Om t \sin \ep 
\nonumber\\
&&
\pt{space}+\sb_B [\cos^2 \Om t \sin^2 \ep -\sin^2 \Om t]),
\label{delps:LV}
\eea
where the combinations of coefficients $\sb_A$ and $\sb_B$ 
are given by
\bea
\sb_A &=& \sin^2 \et \sb_{YY}+ \cos^2 \et \sb_{ZZ}
-\sb_{XX}-\sin 2\et \sb_{YZ},
\nonumber\\
\sb_B &=& \sin \et \sb_{XY}- \cos \et \sb_{XZ}.
\label{sAsB}
\eea

\begin{figure}[h]
\begin{center}
\epsfig{figure=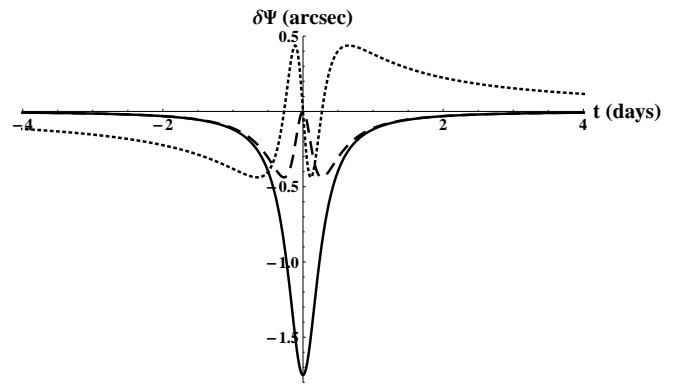,width=1.0\hsize}
\caption{The behavior of the deflection $\de \Ps$ 
near conjunction plotted as function of time $t$ 
around the summer solstice as the Sun moves 
across the field of view.
The solid curve is the deflection from GR, 
the dashed curve is the deflection amplitude 
from the coefficients $\sb_A$, 
and the dotted curve is the deflection amplitude from
the coefficients $\sb_B$. 
\label{conjplot}}
\end{center}
\end{figure}

The result from GR in \rf{delps:GR} is plotted in 
Fig.\ \ref{conjplot} near the time of conjunction ($t=0$)
using the values $\et=23.4^\circ$ and $\ep=0.27^\circ$ (grazing limit).
This deflection is peaked and is symmetrical around $t=0$.
For the Sun as the deflecting body, 
the peak value of $1.75"$ is well known.
Note that the sign of the GR deflection is negative.
This is consistent with the (outward) apparent deflection depicted in 
Fig.\ \ref{fig1}, 
since $\de \Ps_{\rm GR}=\Ps-\Ps_0$ is the difference in 
the observed angle from the unperturbed, 
or zeroth-order angle.

The two types of Lorentz-violating signals in \rf{delps:LV} 
are also plotted in Fig.\ \ref{conjplot} near the time of 
conjunction using the same assumptions on $\et$ and $\ep$.
For these curves we plot amplitudes, or ``partials'', for each 
of the coefficient combinations  $\sb_A$ and $\sb_B$.
It is evident that these signals are qualitatively different 
from the GR case.
The amplitude for $\sb_A$ displays symmetrical behavior
around $t=0$ while the amplitude for $\sb_B$ is
antisymmetric around $t=0$.  
Both of these signals display mildly 
oscillatory behavior near $t=0$ that could potentially
be useful for analysis.

If measurements were obtained for differing orientations of the 
observer, 
the signals for Lorentz violation would involve coefficient
combinations distinct from those in Eq.\ \rf{sAsB}.
As an example of this orientation dependence, 
suppose instead that the (near-conjunction) observations 
took place with the Earth near the vernal equinox, 
when the Sun is along the positive $X$ axis of the chosen 
coordinate system.
The GR result is very similar to 
Eq.\ \rf{delps:GR} in this case, 
but the Lorentz-violating piece $\de \Ps_{\rm LV}$ differs 
in its details.
Specifically, 
for this configuration we find
\bea
\de \Ps_{\rm LV} &\approx&  \fr {GM_{\odot}}{R} 
\fr {\cos \et \sin \Om T}{(1-\cos \Om T \cos \ep
-\sin \et \sin \ep \sin \Om T)^2}
\nonumber\\
&&
\times [(\sb_{YY}-\sb_{ZZ}) (\frac 12 \sin 2\et \sin^2 \Om T 
\nonumber\\
&&
\pt{\times \, \sb_{YY}-\sb_{ZZ}} 
-\cos \et \sin \Om T \cos \Om T \sin \ep)
\nonumber\\
&&
+\sb_{YZ} (\cos^2 \Om T \sin^2 \ep - \cos 2\et \sin^2 \Om T
\nonumber\\
&&
\pt{space}-2\sin \et \sin \ep \sin \Om T \cos \Om T)],
\label{delps:LV2}
\eea
where $T$ is measured from the conjunction time $T=0$.
It is evident from this result that the signal 
depends on the coefficients combinations $\sb_{YY}-\sb_{ZZ}$
and $\sb_{YZ}$, 
rather than those given in Eqs.\ \rf{sAsB}.

\section{Summary and estimates}
\label{summary and estimates}

In this work we have identified the 
dominant signals for local Lorentz violation in light-bending observations.
A general formula making use of euclidean arclength
that is valid for the post-newtonian limit
of any stationary metric was established in \rf{defl:1}.
Working within the SME effective field theory framework,
we applied this formula 
to the deflection of a light ray from a straight line path 
in the grazing limit in Eq.\ \rf{defl:2}, 
and more accurately in Eq.\ \rf{defl:3}.
The results display anisotropic behavior
of light bending controlled by coefficients
for Lorentz violation,
as demonstrated in Fig.\ \ref{fig2}. 

In the latter part of this work, 
we calculate a more practical formula for the
change in the measured angle $\Ps$ between two stars 
in Eq.\ \rf{delps:2}.
We describe generally how this result can be expressed in 
terms of astronomical quantities suitable 
for data analysis.
The approximate behavior of the deflection 
angle shift $\de \Ps$ on the coefficients for Lorentz violation 
was elucidated with a specific near-conjunction example.
We compare this unconventional 
behavior with the standard behavior predicted from GR
(see Fig.\ \ref{conjplot}).

It would be of interest 
to perform a rigorous analysis of potential sensitivities 
for future missions, 
perhaps involving detailed simulations. 
Such simulations have already been performed for the GR 
and PPN case for the 
planned 
Gaia mission \cite{gaia2, gaia3}.
The starting point for 
such simulations is typically
the relative deflection between two stars.
We provide this expression for the 
SME in Eqs.\ \rf{cosps:2} and \rf{delps:2}
of this work.
A numerical least squares estimation could
be attempted using the partial derivatives
of $\Ps$ with respect to the coefficients
for Lorentz violation $\sb_\mn$, 
$\aes_\mu$, 
and $\cs_{00}$, 
along with other relevant parameters.

\begin{table}[h]
\begin{center}
\begin{tabular}{|l|c|c|c|c|}
\hline
Observatory & $\sb'_{TT}$ 
& $\sb_{TJ}+(\al/M)\aes_J$ & $\sb_{JK}$ & Ref. \\
\hline
\hline
LATOR & $10^{-8}$ & $10^{-8}$ 
& $10^{-7}$ & \cite{lator} \\
\hline
Gaia & $10^{-6}$ & $10^{-6}$ 
& $10^{-5}$ & \cite{gaia2} \\
\hline
Hipparcos & $10^{-3}$ & $10^{-3}$ 
& $10^{-2}$ & \cite{hipp} \\
\hline
Optical & $10^{-1}$ & $1$ & $1$
& \cite{optical} \\
\hline
\end{tabular}
\caption{\label{estimates}
Crude estimates of sensitivities of 
current and future light-bending tests
to various combinations of coefficients
for Lorentz violation.
The shorthand $\sb'_{TT}$ is defined by
$\sb'_{TT}=\sb_{TT}+ \cs_{TT}+\frac {\al}{M}\aes_T$.}
\end{center}
\end{table}

Though it is beyond the scope of this work
to perform detailed simulations for future missions
or analysis of available data from current and 
past observations, 
we provide in Table \ref{estimates} 
order of magnitude estimates 
of sensitivities to 
different combinations of coefficients.
These estimates are based on existing 
constraints on deviations from GR in 
light-bending tests or projected measurement accuracies.
We include the proposed Laser Astrometric Test of Relativity 
(LATOR) \cite{lator},
the planned Gaia mission \cite{gaia1}, 
the past Hipparcos mission \cite{hipp}, 
and past ground-based optical observations \cite{optical}.
This is not a comprehensive list and other dedicated
light-bending tests may also be of interest \cite{mgai}.

The estimates in Table
\ref{estimates} can be contrasted with existing constraints
on coefficients for Lorentz violation.
The rotational scalar combination $\sb_{TT}$
has not been formally constrained by rigorous data analysis.  
Care must be taken in light-bending tests since
this coefficients also appears in the Newtonian
force law multiplying the combination $GM$.
Therefore $\sb_{TT}$ is expected to be correlated with orbital tests.
Combining light-bending results with results from orbital tests
could yield the measurements of $\sb_{TT}$ at the levels
indicated in the table.
Similar considerations hold for the scalar matter coefficients
$\aes_T$ and $\cs_{TT}$.
Details on this type of comparative measurement
can be found in Refs.\ \cite{qgrav,tkgrav}.
Note that the matter coefficients can in principle
be separated from $\sb_{TT}$ by using deflecting
bodies of differing composition.
Alternatively, 
one can combine results from light-bending observations
with current constraints on the matter sector coefficients
from earth laboratory experiments \cite{atom2}.

The coefficient combination $\sb_{TJ}+(\al/M)\aes_J$
is also of primary interest for light-bending tests.
The three coefficients $\sb_{TJ}$ are currently
constrained at the $10^{-5}-10^{-6}$ level 
from recent lunar laser ranging and atom interferometry 
tests \cite{tables}.  
For a source body composed of ordinary matter, 
the source combination $\al \aes_J$ depends on the coefficients
for the electron $e$, 
proton $p$, 
and neutron $n$.
Specifically for the Sun as the source body, 
we have 
$\al \aes_J/M \approx 0.5 \,{\rm GeV}^{-1} \al [ (\bar{a}_{\rm eff}^e)_J
+(\bar{a}_{\rm eff}^p)_J+(\bar{a}_{\rm eff}^n)_J]$ \cite{tkgrav}. 
Future missions, 
such as Gaia or LATOR, 
could tighten the constraints on $\sb_{TJ}$ 
and perform the first analysis of astrophysical measurements 
of the matter sector coefficients $(\bar{a}_{\rm eff}^e)_J$,
$(\bar{a}_{\rm eff}^p)_J$,
and $(\bar{a}_{\rm eff}^n)_J$ at the competitive $10^{-6}\, {\rm GeV}$ level
or better.

Other related tests are also of interest.
This includes tests involving the classic time-delay effect \cite{td}.
As mentioned previously in this work, 
modifications to the time delay formula 
arise from local Lorentz violation
and have been analyzed in Refs.\ \cite{qbkgrav, tkgrav}.
Note that the signals for Lorentz violation 
in the time-delay effect and light bending
differ in their dependence on coefficients for Lorentz violation, 
as discussed in Sec.\ \ref{light deflection: grazing case}.
It is therefore of interest to consider all such tests, 
as they could be used to place independent constraints
on Lorentz violation.

Finally we note that we have not treated 
here the broad subject of gravitational lensing.
The deflection angle formulas calculated in this work
could form a starting point for analysis \cite{bt10}.
A comprehensive investigation of the effects of local Lorentz violation
on weak and strong gravitational lensing 
would be of definite interest but lies
beyond the scope of this work \cite{topdef}.


\end{document}